\documentclass[12pt]{article}

\usepackage{amsfonts}
\usepackage{amsmath}
\usepackage{amsthm}
\usepackage{cite}
\usepackage{epsfig}
\usepackage{latexsym}
\usepackage{paralist}
\usepackage{fancyhdr}
\usepackage{graphicx}
\numberwithin{equation}{section}
\usepackage[vcentermath]{youngtab}
\usepackage{young}
\usepackage{ytableau}
\usepackage{etex}
\usepackage{braket}
\usepackage{float}
\usepackage{extarrows}
\usepackage{autobreak}

\usepackage{pict2e}   
\usepackage{animate}  

\usepackage{multirow}
\usepackage{bigdelim}
\usepackage{fancybox}
\usepackage{cals}

\def\be{\begin{equation}}
\def\ee{\end{equation}}
\def\bea{\begin{eqnarray}}
\def\eea{\end{eqnarray}}

\renewcommand{\thefootnote}{\fnsymbol{footnote}}

\begin{document}

\hfuzz=100pt
\title{{\Large \bf{``Chiral'' and ``Non-chiral'' 3d Seiberg duality} }}
\date{}
\author{ Keita Nii$^a$\footnote{nii@itp.unibe.ch}
}
\date{\today}

\maketitle

\thispagestyle{fancy}
\cfoot{}
\renewcommand{\headrulewidth}{0.0pt}

\vspace*{-1cm}
\begin{center}
$^{a}${{\it Albert Einstein Center for Fundamental Physics }}
\\{{\it Institute for Theoretical Physics
}}
\\ {{\it University of Bern}}  
\\{{\it  Sidlerstrasse 5, CH-3012 Bern, Switzerland}}

\end{center}

\begin{abstract}
We propose a Seiberg duality for a 3d $\mathcal{N}=2$ $Spin(7)$ gauge theory with $F$ spinor matters. For $F \ge 6$, the theory allows a magnetic dual description with an $SU(F-4)$ gauge group. The matter content on the magnetic side  is ``chiral'' and the duality connects ``chiral'' and ``non-chiral'' 3d gauge theories. As a corollary, we can construct a Seiberg duality for a 3d $\mathcal{N}=2$ $G_2$ gauge theory with fundamental matters.
\end{abstract}

\renewcommand{\thefootnote}{\arabic{footnote}}
\setcounter{footnote}{0}

\newpage
\tableofcontents 

\newpage

\section{Introduction}

Supersymmetric gauge theories exhibit various low-energy phases depending on the number of dynamical quarks \cite{Seiberg:1994bz}. For small flavors, strongly-coupled gauge dynamics typically breaks supersymmetry and confines the quarks into mesons or baryons. For more flavors, supersymmetric theories flow to a non-abelian Coulomb phase and there are two equivalent ways of describing the low-energy dynamics \cite{Seiberg:1994pq}, which is called ``Seiberg duality.'' After the first Seiberg duality was proposed in a 4d $\mathcal{N}=1$ supersymmetric $SU(N)$ gauge theory with fundamental flavors, this was immediately generalized to various gauge groups and more complicated matters \cite{Intriligator:1995id, Intriligator:1995ne, Kutasov:1995ve, Kutasov:1995np, Intriligator:1995ax}. The similar dualities, which are sometimes called Seiberg-like dualities, were also constructed in three-dimensional spacetime.  

Although a lot of 3d Seiberg dualities have been proposed \cite{Karch:1997ux, Aharony:1997gp, Giveon:2008zn, Niarchos:2008jb, Aharony:2013dha, Aharony:2013kma, Benini:2011mf, Kapustin:2011vz}, its variety of the duality is still limited compared to the 4d dualities. This is because in 3d, there are additional flat directions (moduli spaces) called Coulomb branch from the vector superfield and because the lack of the understanding of its quantum structure makes the construction of the 3d dualities difficult compared to the 4d examples. 
In this paper, we propose the 3d Seiberg duality for the 3d $\mathcal{N}=2$ $Spin(7)$ gauge theory with $F$ spinor matters. The corresponding 4d duality was proposed in \cite{Pouliot:1995zc} and generalized to the $Spin(N)$ cases \cite{Pouliot:1995sk, Cho:1997kr, Pouliot:1996zh, Kawano:1996bd, Berkooz:1997bb, Kawano:2007rz, Kawano:2005nc, Cho:1997sa, Maru:1998hp}. We here claim that a similar duality holds with a small modification of the superpotential. As a by-product of the $Spin(7)$ duality, we can also obtain the duality for the 3d $\mathcal{N}=2$ $G_2$ gauge theory with $F$ fundamental matters. These dualities connect the ``chiral'' and ``non-chiral'' gauge theories from a four-dimensional point of view.

The rest of this paper is organized as follows. 
In Section 2, we study the low-energy dynamics of the 3d $\mathcal{N}=2$ $Spin(7)$ gauge theory with spinor matters, which becomes an electric description of the proposed duality. This section is mostly a review of \cite{Nii:2018tnd}. 
In Section 3, we propose a chiral magnetic description dual to the theory in Section 2 and give various tests of the proposed ``chiral''-``non-chiral'' duality. 
In Section 4, we swap the roles of the electric and magnetic theories. We take the ``chiral'' theory as an electric description and propose a Kutasov-type duality. 
In Section 5, we propose the Seiberg duality for the 3d $\mathcal{N}=2$ $G_2$ gauge theory with fundamental matters. This duality will be derived from the $Spin(7)$ duality via a certain deformation.

\section{3d $\mathcal{N}=2$ $Spin(7)$ gauge theory}

We start with the analysis of the Coulomb branch in the 3d $\mathcal{N}=2$ $Spin(7)$ gauge theory with $F_v$ vector matters and $F_s$ spinor matters \cite{Nii:2018tnd, Nii:2018wwj}. The Coulomb branch is a flat direction spanned by an adjoint scalar in the vector superfield. When the adjoint scalar obtains a non-zero expectation value, the gauge group is spontaneously broken to some subgroup including $U(1)$ factors. Owing to these compact $U(1)$ subgroups, the theory admits monopole configurations which generate a non-perturbative superpotential \cite{Affleck:1982as, Aharony:1997bx, deBoer:1997kr}. This drastically changes the classical picture of the Coulomb branch and a few directions of the Coulomb branch can be quantum-mechanically stable. The quantum Coulomb branch for the $Spin(N)$ theory was studied in \cite{Aharony:2011ci, Aharony:2013kma, Nii:2018tnd, Nii:2018wwj}.

The first Coulomb branch $Y$ was studied in \cite{Aharony:2011ci, Aharony:2013kma} for describing the Coulomb branch of the 3d $\mathcal{N}=2$ $SO(N)$ or $O(N)$ gauge theory with vector matters. When $Y$ obtains a non-zero vev, the gauge group is broken into\footnote{For branching rules in this paper, see for example \cite{Slansky:1981yr, Georgi:1982jb}.}
\begin{align}
so(7) & \rightarrow so(5) \times u(1) \\
\mathbf{7} &\rightarrow  \mathbf{5}_{0} +\mathbf{1}_{2}+\mathbf{1}_{-2}  \\
\mathbf{8} & \rightarrow   \mathbf{4}_{1}+\mathbf{4}_{-1}  \\
\mathbf{21} &  \rightarrow  \mathbf{10}_{0}+\mathbf{1}_{0}+\mathbf{5}_{2}+\mathbf{5}_{-2},
\end{align}
where $\mathbf{7}, \mathbf{8}$ and $\mathbf{21}$ represent vector, spinor and adjoint representations, respectively. Along this Coulomb branch, the components charged under the $U(1)$ subgroup become massive. The masses are proportional to their $U(1)$ charges. Therefore, the vector representation reduces to the massless $SO(5)$ vector whereas all the components of the spinor are massive. When the $Spin(7)$ gauge theory includes vector matters, the low-energy effective theory along $Y$ becomes the $Spin(5)$ gauge theory with massless vectors. The vacuum of this low-energy theory is stable and supersymmetric due to these massless dynamical matters \cite{Aharony:2011ci}. On the other hand, if we consider the $Spin(7)$ gauge theory only with spinor matters, the low-energy theory along $Y$ includes the $3d$ $\mathcal{N}=2$ $Spin(5)$ theory without a dynamical quark, which leads to an unstable vacuum and the supersymmetry is lost due to the Affleck-Harvey-Witten superpotential \cite{Affleck:1982as}. As a result, we need to introduce this Coulomb branch operator for the $Spin(7)$ gauge theory with vector matters. In what follows, we will only study the 3d $\mathcal{N}=2$ $Spin(7)$ gauge theory with spinor matters and $Y$ is not necessary.

For the $Spin(7)$ gauge theory with spinor matters, there is an additional Coulomb branch \cite{Nii:2018tnd, Nii:2018wwj}. 
The second Coulomb branch denoted by $Z$ corresponds to the following gauge symmetry breaking
\begin{align}
so(7) & \rightarrow so(3) \times su(2) \times u(1) \\
\mathbf{7} &\rightarrow (\mathbf{3}, \mathbf{1})_{0}+(\mathbf{1}, \mathbf{2})_{\pm 1} \\
\mathbf{8} & \rightarrow   (\mathbf{2}, \mathbf{1})_{\pm 1} +(\mathbf{2}, \mathbf{2})_{0} \\
\mathbf{21} &  \rightarrow (\mathbf{3}, \mathbf{1})_{0} +(\mathbf{1}, \mathbf{3})_{0}+(\mathbf{1}, \mathbf{1})_{0} + (\mathbf{3}, \mathbf{2})_{\pm 1}+(\mathbf{1}, \mathbf{1})_{\pm 2},  
\end{align}
where the Coulomb branch $Z$ is defined from the $U(1)$ subgroup by dualizing the $U(1)$ vector superfield into a chiral superfied.
Since the components charged under the $U(1)$ subgroup have non-zero masses, the vector representation reduces to $(\mathbf{3}, \mathbf{1})_{0}$. When we consider the $Spin(7)$ gauge theory only with vector matters, the low-energy theory along this direction includes a 3d $\mathcal{N}=2$ pure $SU(2)$ gauge theory whose vacuum is runaway and unstable \cite{Affleck:1982as}. Therefore, the $Spin(7)$ gauge theory only with vector matters cannot have this flat direction. For the theory with spinor matters, there are massless components $(\mathbf{2}, \mathbf{2})_{0}$ and the low-energy $SO(3) \times SU(2) \times U(1)$ gauge theory can have a stable and supersymmetric vacuum. In this paper, we will discuss the Seiberg duality of the 3d $\mathcal{N}=2$ $Spin(7)$ gauge theory with $F$ spinor matters. Therefore, we only consider the single Coulomb branch $Z$. Table \ref{Fermionzero} summarizes the numbers of the fermion zero-modes for the monopoles associated with $Y$ and $Z$.

\begin{table}[H]\caption{Fermion zero-modes around the monopoles} 
\begin{center}
\scalebox{1}{
  \begin{tabular}{|c||c|c|c| } \hline
  & adjoint (gaugino) $\mathbf{21}$ & vector $\mathbf{7}$ & spinor $\mathbf{8} $ \\  \hline 
$Y$&$10$&$2F_v$& $4F_s$  \\
$Z$ &$8$&$2F_{v}$& $2F_{s}$  \\ \hline
  \end{tabular}}
  \end{center}\label{Fermionzero}
\end{table}

The electric theory is a 3d $\mathcal{N}=2$ $Spin(7)$ gauge theory with $F$ spinor matters \cite{Nii:2018tnd}. The Higgs branch is described by the meson operator $M:=QQ$ and the baryon operator $B:=Q^4$. The Coulomb branch is described by $Z$. The quantum numbers of the elementary fields and moduli coordinates are summarized in Table \ref{Spin7electric}.

\begin{table}[H]\caption{3d $\mathcal{N}=2$ $Spin(7)$ gauge theory with $F$ spinors} 
\begin{center}
\scalebox{1}{
  \begin{tabular}{|c||c|c|c|c| } \hline
  &$Spin(7)$&$SU(F)$&$U(1)$& $U(1)_R$  \\ \hline
$Q$&$\mathbf{8}$&${\tiny \yng(1)}$&1&$0$ \\   \hline 
$M:=QQ$&1&${\tiny \yng(2)}$&2&$0$  \\
$B:=Q^4$&1&${\tiny \yng(1,1,1,1)}$&4&$0$ \\[8pt]  \hline
$Z$&1&1&$-2F$&$2F-8$  \\  \hline
  \end{tabular}}
  \end{center}\label{Spin7electric}
\end{table}

We briefly summarize the low-energy dynamics for the cases with small flavors $F \le 5$ \cite{Nii:2018tnd}: For $F=5$, the theory is in an s-confinement phase. The low-energy dynamics is described by the gauge-invariant chiral superfields $M,B$ and $Z$ with an effective superpotential
\begin{align}
W_{eff, \, F=5} =Z \left( \det \, M -BMB \right). \label{Wspin7F5}
\end{align}
At the origin of the moduli space, the confinement phase without symmetry breaking is realized, which is called s-confinement. 
For $F=4$, the theory is again in a confining phase described by $M,B$ and $Z$ with a single quantum constraint 
\begin{align}
W_{eff, \, F=4} =\lambda \left[ Z (\det \, M -B^2) -1 \right],
\end{align}
where $\lambda$ is a Lagrange multiplier. One should notice that the origin of the moduli space is eliminated by this constraint. Therefore, the confinement for $F=4$ necessarily induces some symmetry breaking. For $F \le 3$, the moduli space is described by $M$ and $Z$. We can write down an effective superpotential
\begin{align}
W_{eff, \, F\le 3} = \left(  \frac{1}{Z \det \, M}  \right)^{\frac{1}{4-F}}
\end{align}
and there is no stable supersymmetric vacuum for $F \le 3$. For $F \ge 6$, we can anticipate that the theory is in a non-abelian Coulomb phase and that there is a magnetic dual description. This expectation is plausible as follows: The 3d $\mathcal{N}=2$ $Spin(7)$ gauge theory with $F$ spinors flows to the 3d $\mathcal{N}=2$ $SU(3)$ gauge theory with $F-2$ fundamental flavors via a deformation with $\mathrm{rank}\, \braket{M} = 2$. The low-energy $SU(3)$ gauge theory exhibits a non-abelian Coulomb phase for $F \ge 6$ \cite{Aharony:1997bx, Aharony:2013dha} and there are additional massless degrees of freedom at the origin of the moduli space. Therefore, it is natural to think that the $Spin(7)$ theory with $F \ge 6$ also exhibits an interacting non-abelian Coulomb phase. In the next section, we will propose a magnetic description dual to Table \ref{Spin7electric} by imitating the 4d construction of the $Spin(7)$ duality \cite{Pouliot:1995zc}.

\section{The $SU(F-4)$ magnetic dual}

Now, we propose a magnetic description dual to Table \ref{Spin7electric}, which is very similar to the corresponding 4d duality proposed by Pouliot \cite{Pouliot:1995zc}. The magnetic side becomes a 3d $\mathcal{N}=2$ $SU(F-4)$ gauge theory with $F$ anti-fundamental matters $\tilde{q}$, a symmetric tensor $s$ and a meson singlet $M$. The meson field $M$ is identified with the electric meson $QQ$. Table \ref{Spin7magnetic} summarizes the quantum numbers of these elementary fields.

Let us study the Coulomb branch of the magnetic theory. When the bare Coulomb branch operator $Y_{SU(F-6)}^{bare}$ obtains a non-zero expectation value, the gauge group is spontaneously broken to
\begin{align}
SU(F-4) & \rightarrow SU(F-6) \times U(1)_1 \times U(1)_2  \\
{\tiny \overline{ \yng(1)}} & \rightarrow  {\tiny \overline{ \yng(1)}}_{\, 0,2}+1_{-1,-(F-6)} +1_{1,-(F-6)} \\
{\tiny  \yng(2)} & \rightarrow {\tiny  \yng(2)}_{\, 0,-4} +{\tiny  \yng(1)}_{\, 1, F-8} +{\tiny  \yng(1)}_{\, -1,F-8} +1_{2,2F-12} +1_{-2,2F-12}+1_{0,2F-12}.
\end{align}
The Coulomb branch is associated with the $U(1)_1$ subgroup and its coordinate $Y_{SU(F-6)}^{bare}$ is constructed from the $U(1)_1$ vector superfield. Along this breaking, the components charged under the $U(1)_1$ symmetry are massive and integrated out.  
Since the matter content is ``chiral,''  the fermion one-loop diagrams with these massive components induce the mixed Chern-Simons term between the $U(1)_1$ and $U(1)_2$ groups. This results in a non-zero $U(1)_2$ charge of $Y_{SU(F-6)}^{bare}$ \cite{Intriligator:2013lca}. The quantum numbers of the bare Coulomb branch $Y_{SU(F-6)}^{bare}$ is listed in Table \ref{Spin7magnetic}. In order to parametrize the Coulomb branch, we need to define a so-called dressed monopole operator
\begin{align}
Y^{dressed}:= Y^{bare}_{SU(F-4)} \left(  {\tiny  \yng(2)}_{\, 0,-4} \right)^{F-6}  \sim  Y^{bare}_{SU(F-4)} s^{F-6},
\end{align}
where the color indices of $s^{F-6}$ are contracted with an epsilon tensor of the $SU(F-6)$ subgroup. The $U(1)_2$ charge of the bare operator is correctly canceled as it should be.

The magnetic theory has a tree-level superpotential
\begin{align}
W_{mag}= Ms \tilde{q} \tilde{q} +Y^{dressed}, \label{SUmagW}
\end{align}
which is consistent with all the global symmetries in Table \ref{Spin7magnetic}. Several comments about this superpotential are in order: As opposed to the 4d case, we need not introduce a term proportional to $\det \, s$ into the superpotential \eqref{SUmagW}. The magnetic Coulomb branch $Y^{dressed}$ is excluded from the chiral ring elements via the superpotential \eqref{SUmagW}. This monopole superpotential is absent in the corresponding 4d duality but reminiscent of the 4d superpotential since the 4d theory includes the term proportional to $ \det \, s\sim  s^{F-4}$ instead of $Y^{dressed}\sim  Y^{bare}_{SU(F-4)} s^{F-6}$. From this point of view, the superpotential \eqref{SUmagW} is similar to the 4d one. Notice that the assignment of the global $U(1)$ charge on the magnetic elementary fields are completely fixed by requiring that the baryon matching $B\sim \tilde{q}^{F-4}$ and the consistency of the first term $ Ms \tilde{q} \tilde{q}$ in the superpotential \eqref{SUmagW}. The availability of the monopole potential and the other operator matching $Z \sim \det \, s$ give us a non-trivial consistency check of the proposed duality.

\begin{table}[H]\caption{The $SU(F-4)$ magnetic dual of Table \ref{Spin7electric}} 
\begin{center}
\scalebox{1}{
  \begin{tabular}{|c||c|c|c|c| } \hline
  &$SU(F-4)$&$SU(F)$&$U(1)$& $U(1)_R$  \\ \hline
$\tilde{q}$&${\tiny \overline{ \yng(1)}}$&${\tiny \overline{\yng(1)}}$&$\frac{4}{F-4}$&$0$ \\  
$s$&${\tiny \yng(2)}$&1&$\frac{-2F}{F-4}$&$2$ \\ 
$M$&1&${\tiny \yng(2)}$&2&$0$  \\   \hline 
$B\sim \tilde{q}^{F-4}$&1& ${\tiny \yng(1,1,1,1)}$&4&$0$ \\[7pt]  
$Z \sim \det\,s$&1&1&$-2F$&$2F-8$  \\  \hline 
$Y^{bare}_{SU(F-4)}$&$U(1)_2$ charge: $4(F-6)$&1&$2F-\frac{4F}{F-4}$&$-2F+14$  \\
$Y^{dressed}:= Y^{bare}_{SU(F-4)} s^{F-6}$&1&1&0&$2$  \\ \hline
  \end{tabular}}
  \end{center}\label{Spin7magnetic}
\end{table}

First, we consider a complex mass deformation. This is completely the same as the 4d argument \cite{Pouliot:1995zc}. Let us introduce a complex mass to the last flavor of the spinor matters $W=m Q_FQ_F =m M_{FF}$. By taking a low-energy limit at the origin of the moduli space, the electric theory flows to the $Spin(7)$ theory with $F-1$ spinor matters. On the magnetic side, the deformation $W=m M_{FF}$ corresponds to the higgsing $\braket{s \tilde{q}^{F} \tilde{q}^F} = -m$ which breaks the gauge group into $SU(F-5)$. At the low-energy limit, we obtain the $SU(F-5)$ gauge theory with $F-1$ anti-fundamental matters and a symmetric tensor. In this way, the mass deformation preserves the duality with reduction $F \rightarrow F-1$. 

Let us further test the validity of our duality. For $F=5$, the electric $Spin(7)$ gauge theory exhibits the s-confinement phase \eqref{Wspin7F5}. On the magnetic side, the gauge group is vanishing. The theory is described by the gauge-singlet chiral superfields $\tilde{q},s$ and $M$ which are identified with the electric moduli coordinates $B, Z$ and $M$, respectively. The magnetic superpotential becomes
\begin{align}
W_{mag} =Ms \tilde{q} \tilde{q} =ZBMB,
\end{align}
which reproduces a part of the s-confinement superpotential \eqref{Wspin7F5}. We expect that the missing term $Z \det \, M$ is non-perturbatively generated, which is consistent with the global symmetries in Table \ref{Spin7magnetic}.

Finally, we can compute the superconformal indices \cite{Bhattacharya:2008bja, Kim:2009wb, Imamura:2011su, Kapustin:2011jm} from the electric and magnetic descriptions and find a beautiful agreement. We here focus on the case with $F=6$ where the magnetic side becomes the 3d $\mathcal{N}=2$ $SU(2)$ gauge theory with six doublets and an adjoint matter. The magnetic Coulomb branch need not be dressed in the $SU(2)$ case. We computed the superconformal indices from both the electric and magnetic sides and the results are given by 

\footnotesize
\begin{align}
I&=1+21 u^2 x^{1/2}+\left(\frac{1}{u^{12}}+246 u^4\right) x+\left(\frac{21}{u^{10}}+2086 u^6\right) x^{3/2}+\left(\frac{1}{u^{24}}+14196 u^8+\frac{246}{u^8}-36\right) x^2  \nonumber \\
&+\left(\frac{21}{u^{22}}+81879 u^{10}+\frac{2051}{u^6}-840 u^2\right) x^{5/2}+\left(\frac{1}{u^{36}}+\frac{246}{u^{20}}+413924 u^{12}-\frac{35}{u^{12}}-10500 u^4+\frac{13377}{u^4}\right) x^3+\cdots,
\end{align}
\normalsize

\noindent where we set the $R$-charge of $Q$ to be $r_Q=\frac{1}{4}$ for simplicity. The fugacity $u$ is associated with the global $U(1)$ symmetry and $x$ counts the weight plus the third component of spin.  The second term $21 u^2 x^{1/2}$ corresponds to the meson operator $M:=QQ$. The baryon operator $B:=Q^4$ is represented as $15 u^4 x$ and the remaining $231 u^4x$ is $M^2$. The higher order terms are symmetric products of these operators and fermion contributions. We checked the agreement of the electric and magnetic indices up to $O(x^3)$. We also computed the superconformal indices for the case with $F=7$ and observed the agreement up to $O(x^3)$.

\section{Kutasov-type duality}
In this section, we regard the $SU(N)$ gauge theory with a symmetric tensor as an electric description and propose a Kutasov-type\footnote{``Kutasov-type'' here means Seiberg dualities where the electric theory has some tree-level superpotential which truncates the chiral ring and simplifies the construction of the dualities \cite{Kutasov:1995ve, Kutasov:1995np}.} duality according to Pouliot's argument \cite{Pouliot:1995zc}. The electric side becomes a 3d $\mathcal{N}=2$ $SU(N)$ gauge theory with $N+4$ fundamental matters $Q$ and a symmetric-bar tensor $\bar{S}$, where we shifted $F \rightarrow N+4$ for simplicity. The electric theory includes the following superpotential
\begin{align}
W_{ele} = Y^{dressed}, \label{WKutasovele}
\end{align}
where $ Y^{dressed}$ is a dressed Coulomb branch operator and its precise form will be defined below. Notice that Pouliot's 4d duality \cite{Pouliot:1995zc} has a different superpotential $W^{4d}_{ele} = \det \, \bar{S}$ whereas the 3d case includes a monopole superpotential \eqref{WKutasovele}. Since there is no superpotential for the Higgs branch coordinates, the Higgs branch operators are not truncated at all and described by
\begin{align}
M:= \bar{S}QQ,~~~~B:=Q^N,~~~~\det \, \bar{S}.
\end{align}
The analysis of the Coulomb branch is very similar to the previous section. 
The bare Coulomb branch denoted by $Y^{bare}_{SU(N-2)}$ corresponds to the gauge symmetry breaking
\begin{align}
SU(N) & \rightarrow SU(N-2) \times U(1)_1 \times U(1)_2  \\
{\tiny \yng(1)} & \rightarrow  {\tiny \yng(1)}_{\, 0,-2}+1_{1,N-2} +1_{-1,-(N-2)} \\
{\tiny  \overline{\yng(2)}} & \rightarrow {\tiny  \overline{\yng(2)}}_{\, 0,4} +{\tiny  \overline{\yng(1)}}_{\, -1, -(N-4)} +{\tiny  \overline{\yng(1)}}_{\, 1,-(N-4)} +1_{-2,-(2N-4)} +1_{2,-(2N-4)}+1_{0,-(2N-4)},
\end{align}
where the Coulomb branch corresponds to the vector superfield of the unbroken $U(1)_1$ subgroup.
Along this flat direction, the mixed Chern-Simons term between the $U(1)_1$ and $U(1)_2$ symmetries is generated. This turns on a non-trivial $U(1)_2$ charge to $Y^{bare}_{SU(N-2)}$ \cite{Intriligator:2013lca}. 
The $U(1)_2$ charge of $Y^{bare}_{SU(N-2)}$ is $-4(N-2)$ and the dressed operator is defined by 
\begin{align}
Y^{dressed} := Y^{bare}_{SU(N-2)}  \left( {\tiny  \overline{\yng(2)}}_{\, 0,4} \right)^{N-2} \sim  Y^{bare}_{SU(N-2)}  \bar{S}^{N-2},
\end{align}
where the color indices of $\bar{S}^{N-2}$ are contracted by an epsilon tensor of the unbroken $SU(N-2)$ subgroup. This dressed operator is eliminated from the chiral ring elements due to the superpotential \eqref{WKutasovele}. Table \ref{Kutasovele} summarizes the quantum numbers of the elementary fields and moduli coordinates. One should notice that the $U(1)$ charge assignment is completely fixed by the monopole superpotential \eqref{WKutasovele}.

\begin{table}[H]\caption{3d $\mathcal{N}=2$ $SU(N)$ gauge theory with $(N+4)\, {\tiny \protect\yng(1)} + {\tiny \overline{\protect\yng(2)}} $ } 
\begin{center}
\scalebox{1}{
  \begin{tabular}{|c||c|c|c|c| } \hline
  &\small $SU(N)$&\small $SU(N+4)$&\small $U(1)$& \small $U(1)_R$  \\ \hline  
$Q$&${\tiny \yng(1)}$&${\tiny \yng(1)}$&$1$&0\\
$\bar{S}$&${\tiny \overline{ \yng(2)}}$&1&$-\frac{N+4}{2}$&$2$  \\  \hline
$M:= \bar{S}QQ$&1&${\tiny \yng(2)}$&$-\frac{N}{2}$&$2$  \\
$B:=Q^N$&1&${\tiny \overline{ \yng(1,1,1,1)}}$&$N$&$0$  \\[7pt]
$\det \, \bar{S} $&1&1&$-\frac{N(N+4)}{2}$&$2N$  \\ \hline 
$Y^{bare}_{SU(N-2)}$&\scriptsize $U(1)_2$ charge: $-4(N-2)$&1&\scriptsize $-(N+4)+\frac{N(N+4)}{2}$&$6-2N$  \\
$\small Y^{dressed}:=Y^{bare}_{SU(N-2)} \bar{S}^{N-2}$&1&1&0&2  \\ \hline
  \end{tabular}}
  \end{center}\label{Kutasovele}
\end{table}

The magnetic side becomes a 3d $\mathcal{N}=2$ $Spin(7)$ gauge theory with $N+4$ spinor matters $q$ and a meson singlet $M$ which is identified with the electric meson $\bar{S}QQ$. The magnetic theory includes a tree-level superpotential
\begin{align}
W_{mag} =Mqq.  \label{Wkutasovmag}
\end{align}
Table \ref{Kutasovmag} summarizes the quantum numbers of the elementary fields in the magnetic $Spin(7)$ gauge theory. Notice that the assignment of the global $U(1)$ charge in Table \ref{Kutasovmag} is fixed by the meson matching $M\sim \bar{S}QQ$ and the superpotential \eqref{Wkutasovmag}. Therefore, the matching of the other operators becomes a non-trivial test of the duality as we will see below. 
 
\begin{table}[H]\caption{The $Spin(7)$ magnetic dual description of Table \ref{Kutasovele}} 
\begin{center}
\scalebox{1}{
  \begin{tabular}{|c||c|c|c|c| } \hline
  &$Spin(7)$&$SU(N+4)$&$U(1)$& $U(1)_R$  \\ \hline
$q$&$\mathbf{8}$&${\tiny \overline{ \yng(1)}}$&$\frac{N}{4}$&$0$ \\   
$M$&1&${\tiny \yng(2)}$&$-\frac{N}{2}$&$2$  \\    \hline
$B \sim q^4$&1&${\tiny \overline{\yng(1,1,1,1)}}$&$N$&$0$ \\[8pt]  \hline
$Z$&1&1&$-\frac{N(N+4)}{2}$&$2N$  \\  \hline
  \end{tabular}}
  \end{center}\label{Kutasovmag}
\end{table}

Due to the superpotential \eqref{Wkutasovmag}, the magnetic meson $qq$ is eliminated from the moduli space of the magnetic theory. The matching of the gauge invariant operators are easily obtained by comparing Table \ref{Kutasovele} with Table \ref{Kutasovmag}:
\begin{gather*}
Q^N \sim q^4,~~~~\bar{S}QQ \sim M,~~~~\det \, \bar{S} \sim Z
\end{gather*}
Notice that in the corresponding 4d duality \cite{Pouliot:1995zc}, the composite operator $\det \, \bar{S} $ was excluded due to a tree-level superpotential $W^{4d}_{ele} = \det \, \bar{S}$ while it is here mapped to the Coulomb branch operator $Z$.

Let us compare the flat directions of the electric and magnetic theories. In the 4d case where a different superpotential $W_{ele}^{4d} =\det \, \bar{S}$ is introduced,  the $F$-flatness condition imposes $\mathrm{rank} \, \braket{M}=\braket{QQ\bar{S}} \le N-2$. On the other hand, in 3d, $\bar{S}$ is not constrained by the superpotential \eqref{WKutasovele}. Therefore, $\bar{S}$ (and $M:=QQ \bar{S}$ as well) can have non-zero expectation values such that $\mathrm{rank} \, \braket{M} \le N$. We can see that the meson singlet $M$ on the magnetic side is in the same situation: If the singlet $M$ had a non-zero vev with $\mathrm{rank} \, \braket {M} \ge N+1$, the magnetic theory flows to a 3d $\mathcal{N}=2$ $Spin(7)$ theory with less than four spinor matters. As we explained in Section 2, such a theory exhibits a runaway superpotential and loses all the supersymmetric vacua. Therefore, the rank of the meson singlet on the magnetic side should be less than or equal to $N$. In this way, the electric and magnetic theories have the same mesonic flat directions parametrized by $M=\bar{S}QQ$.

As a further consistency check, let us consider a trivial case with $N=1$, where $SU(N)$ is vanishing. The electric side becomes a non-gauge theory with two gauge-singlet chiral superfields $Q$ and $\bar{S}$. These are free fields. When $N=1$, the magnetic side becomes a 3d $\mathcal{N}=2$ $Spin(7)$ gauge theory with five spinors. As explained in Section 2, this magnetic theory is s-confined and described by the effective superpotential
\begin{align}
W_{mag}^{eff} =MN +Z(\det \, N -BNB),
\end{align}
where we defined a magnetic meson $N:=qq$. The two mesonic fields $M$ and $N$ are massive. By integrating out these massive fields, we are left with the two free fields $B$ and $Z$ which are identified with $Q$ and $\bar{S}$, respectively.

\section{3d $G_2$ Seiberg duality}
In this section, we propose a Seiberg duality for the 3d $\mathcal{N}=2$ $G_2$ gauge theory with $F$ fundamental matters \cite{Nii:2017npz}. The dimension of the fundamental representation in $G_2$ is $\mathbf{7}$. This duality can be derived from the $Spin(7)$ duality with $F+1$ spinor matters, which was studied in Section 2 and Section 3, by introducing a rank-one vev to $M$. On the electric side, the vev for the last flavor $\braket{M_{F+1,F+1}}\neq 0$ breaks the gauge group into $G_2$ and one spinor is eaten via the Higgs mechanism. On the magnetic side (shifting $F \rightarrow F+1$), the gauge group is $SU(F-3)$ and not higgsed. The vev for $\braket{M_{F+1,F+1}}\neq 0$ decomposes $M s \tilde{q} \tilde{q} \rightarrow Ms \tilde{q} \tilde{q} +s \tilde{q}_0 \tilde{q}_0$, where we absorbed the vev of $M_{F+1,F+1}$ into $\tilde{q}_0$'s. In this way, we can derive the 3d duality between the $G_2$ and $SU(F-3)$ gauge theories. In what follows, we will investigate this duality in further detail. 

The electric theory is a 3d $\mathcal{N}=2$ $G_2$ gauge theory with $F$ fundamental matters. The supersymmetry-breaking and confinement phases for $F \le 4$ were investigated in \cite{Nii:2017npz}. Here, we focus on $F \ge 5$. The global symmetry is $SU(F) \times U(1) \times U(1)_R$ where the $U(1)$ symmetry counts the number of quarks. Notice that there is no chiral anomaly in 3d and that the 3d parity anomaly impose no constraint on the matter content. There is no tree-level superpotential on the electric side. The quantum numbers of the elementary fields are summarized in Table \ref{G2electric}.

The Higgs branch, which is identical to the 4d one, is described by the following gauge-invariant composites \cite{Pesando:1995bq, Giddings:1995ns, Pouliot:2001iw}
\begin{align}
M:=QQ,~~~~B:=Q^3,~~~F:=Q^4,
\end{align}
where the color indices are contracted by $\delta^{a_1a_2}, \, f^{a_1a_2a_3}$ and $\epsilon^{a_1 \cdots a_7} f^{a_5 a_6 a_7}$ $(a_i =1,\cdots,7)$. 
Next, we consider the Coulomb branch from the $G_2$ vector superfield. The $G_2$ Coulomb branch denoted by $Z_{SU(2)}$ was studied in \cite{Nii:2017npz, Nii:2019dwi}. When $Z_{SU(2)}$ obtains a non-zero expectation value, the gauge group is spontaneously broken to  
\begin{align}
G_2 & \rightarrow su(2) \times u(1)  \\
\mathbf{7} & \rightarrow   \mathbf{2}_{\pm 1}+ \mathbf{3}_{0}  \\
\mathbf{14} &  \rightarrow    \mathbf{3}_{0}+ \mathbf{1}_{0} + \mathbf{1}_{\pm  2}+ \mathbf{4}_{\pm 1},
\end{align}
where $\mathbf{14}$ is an adjoint representation. Since the fundamental matter reduces to a massless $SU(2)$ triplet along this flat direction, the low-energy $SU(2)$ gauge theory can have a stable supersymmetric vacuum \cite{Affleck:1982as, Aharony:1997bx, deBoer:1997kr}. Therefore, the Coulomb branch labeled by $Z_{SU(2)}$ is quantum-mechanically allowed. For other Coulomb branch directions, there is no massless dynamical quark from the fundamental representation and hence the low-energy $SU(2)$ gauge theory makes its vacuum unstable \cite{Affleck:1982as}. As a result, the quantum Coulomb branch is described by a single coordinate $Z_{SU(2)}$.

\begin{table}[H]\caption{3d $\mathcal{N}=2$ $G_2$ gauge theory with $F$ fundamental matters} 
\begin{center}
\scalebox{1}{
  \begin{tabular}{|c||c|c|c|c| } \hline
  &$G_2$&$SU(F)$&$U(1)$& $U(1)_R$  \\ \hline
$Q$&$\mathbf{7}$&${\tiny \yng(1)}$&1&$0$ \\   \hline 
$M:=QQ$&1&${\tiny \yng(2)}$&2&$0$  \\
$B:=Q^3$&1&${\tiny \yng(1,1,1)}$&3&$0$ \\[6pt]  
$F:=Q^4$&1&${\tiny \yng(1,1,1,1)}$&4&$0$ \\[8pt]  \hline
$Z_{SU(2)}$&1&1&$-2F$&$2F-6$  \\  \hline
  \end{tabular}}
  \end{center}\label{G2electric}
\end{table}

The magnetic description is given by the 3d $\mathcal{N}=2$ $SU(F-3)$ gauge theory with $F+1$ anti-fundamental matters, a symmetric tensor $s$ and a meson singlet $M$. The meson field $M$ is identified with the electric meson $QQ$. The anti-fundamental matters are decomposed into $F$ anti-quarks $\tilde{q}$ and a single one $\tilde{q}_0$ as we explained before. The magnetic theory includes a tree-level superpotential
\begin{align}
W_{mag} =Ms \tilde{q} \tilde{q} +s \tilde{q}_0 \tilde{q}_0 +Y^{dressed},
\end{align}
which distinguishes $\tilde{q}_0$ from $\tilde{q}$. The last term $Y^{dressed}$ is a dressed Coulomb branch operator of the magnetic theory, which will be defined below. The quantum numbers of the magnetic elementary fields are summarized in Table \ref{G2magnetic}. The global $U(1)$ charge assignment on the magnetic side is completely determined only by requiring the baryon matching $B:= Q^3 \sim \tilde{q}^{F-3}$ and the availability of $Ms \tilde{q} \tilde{q} +s \tilde{q}_0 \tilde{q}_0$ in the superpotential. Therefore, it is a non-trivial consistency check that we can have the superpotential $W = Y^{dressed}$ consistent with the symmetries in Table \ref{G2magnetic}. As we will see below, it is the case.

The analysis of the Coulomb branch is the same as the previous sections. The bare Coulomb branch operator denoted by $ Y^{bare}_{SU(F-5)} $ corresponds to the gauge symmetry breaking
\begin{align}
SU(F-3) & \rightarrow SU(F-5) \times U(1)_1 \times U(1)_2  \\
{\tiny \overline{ \yng(1)}} & \rightarrow  {\tiny \overline{ \yng(1)}}_{\, 0,2}+1_{-1,-(F-5)} +1_{1,-(F-5)} \\
{\tiny  \yng(2)} & \rightarrow {\tiny \yng(2)}_{\, 0,-4} +{\tiny \yng(1)}_{\, 1, F-7} +{\tiny  \yng(1)}_{\, -1,F-7} +1_{2,2F-10} +1_{-2,2F-10}+1_{0,2F-10},
\end{align}
where $ Y^{bare}_{SU(F-5)} $ is constructed from the $U(1)_1$ vector superfield. Since the bare operator has a non-zero $U(1)_2$ charge proportional to the effective mixed Chern-Simons term between $U(1)_1$ and $U(1)_2$ \cite{Intriligator:2013lca}, we need to define the dressed monopole operator
\begin{align}
Y^{dressed}:= Y^{bare}_{SU(F-5)} \left(  {\tiny  \yng(2)}_{\, 0,-4} \right)^{F-5}    \sim Y^{bare}_{SU(F-5)} s^{F-5},
\end{align}
where the color indices of $s^{F-5}$ are contracted with an epsilon tensor of the $SU(F-5)$ gauge group. Note that the global $U(1)$ charge is correctly canceled in this combination as in Table \ref{G2magnetic}.

\begin{table}[H]\caption{The $SU(F-3)$ magnetic dual description of Table \ref{G2electric}} 
\begin{center}
\scalebox{1}{
  \begin{tabular}{|c||c|c|c|c| } \hline
  &$SU(F-3)$&$SU(F)$&$U(1)$& $U(1)_R$  \\ \hline  
$\tilde{q}$&${\tiny \overline{ \yng(1)}}$&${\tiny \overline{ \yng(1)}}$&$\frac{3}{F-3}$&0\\
$\tilde{q}$&${\tiny \overline{ \yng(1)}}$&1&$\frac{F}{F-3}$&0  \\
$s$&${\tiny  \yng(2)}$&1&$-\frac{2F}{F-3}$&$2$  \\ 
$M$&1&${\tiny \yng(2)}$&2&$0$  \\  \hline
$B \sim \tilde{q}^{F-3}$&1&${\tiny \yng(1,1,1)}$&3&$0$ \\[6pt]  
$F  \sim \tilde{q}^{F-4} \tilde{q}_0 $&1&${\tiny \yng(1,1,1,1)}$&4&$0$ \\[8pt]  
$Z_{SU(2)} \sim \det \, s$&1&1&$-2F$&$2F-6$  \\ \hline
$Y^{bare}_{SU(F-5)}$&$U(1)_2$ charge: $4(F-5)$&1&$2F-\frac{4F}{F-3}$&$-2F+12$  \\
$Y^{dressed}:=Y^{bare}_{SU(F-5)} s^{F-5}$&1&1&0&2  \\  \hline
  \end{tabular}}
  \end{center}\label{G2magnetic}
\end{table}

The matching of the chiral ring elements under the $G_2$ duality is transparent from Table \ref{G2electric} and Table \ref{G2magnetic}:
\begin{gather*}
QQ  \sim M,~~~~Q^3 \sim \tilde{q}^{F-3} \\
Q^4 \sim \tilde{q}^{F-4} \tilde{q}_0,~~~~Z_{SU(2)} \sim \det \, s
\end{gather*}
Notice that the matching of the quartic baryon $F:=Q^4 \sim \tilde{q}^{F-4} \tilde{q}_0 $ and the Coulomb branch $Z_{SU(2)}  \sim \det \, s$ is non-trivial and serves as a consistency check of our duality.
The magnetic superpotential lifts unnecessary flat directions $s \tilde{q}\tilde{q}, \,  s \tilde{q}_0 \tilde{q}_0$ and $Y^{dressed}$.

We can verify further consistencies of the $G_2$ duality. For $F=4$, the electric theory is in an s-confinement phase \cite{Nii:2017npz}. The confined degrees of freedom are $M, B, F$ and $Z_{SU(2)}$. The effective superpotential for $F=4$ becomes 
\begin{align}
W_{ele}^{eff} = Z_{SU(2)} \left( \det \, M +BMB +F^2\right), \label{WG2F4}
\end{align}
which governs the low-energy dynamics.
On the magnetic side with $F=4$, the gauge group $SU(F-3)$ vanishes. Thus, the theory is described by the gauge-invariant chiral superfields $\tilde{q}, \tilde{q}_0, s$ and $M$. The magnetic superpotential becomes
\begin{align}
W_{mag} &=Ms\tilde{q}\tilde{q}+s \tilde{q} \tilde{q} \nonumber \\
&= Z_{SU(2)} \left( BMB +F^2 \right),
\end{align}
where we used the following operator identification
\begin{align}
B \sim \tilde{q},~~~F \sim \tilde{q}_0,~~~Z_{SU(2)} \sim s,
\end{align}
which is consistent with the global symmetries in Table \ref{G2electric} and Table \ref{G2magnetic}. In this way, we can reproduce a part of the electric superpotential \eqref{WG2F4} except for $Z_{SU(2)}  \det \, M $. We expect that this missing term is dynamically generated on the magnetic side since  $Z_{SU(2)}  \det \, M $ is consistent with the magnetic global symmetries.

As a final consistency check, we compute the superconformal indices \cite{Bhattacharya:2008bja, Kim:2009wb, Imamura:2011su, Kapustin:2011jm} by using the electric and magnetic descriptions. We here focus on the case with $F=5$. We find that the two descriptions give us an identical result: 

\scriptsize
\begin{align}
I &= 1+\left(\frac{1}{t^{10}}+15 t^2\right) x^{2/3}+10 t^3 x+\left(\frac{1}{t^{20}}+\frac{15}{t^8}+125 t^4\right) x^{4/3}+\left(\frac{10}{t^7}+150 t^5\right) x^{5/3} \nonumber \\
&\quad +\left(\frac{1}{t^{30}}+\frac{15}{t^{18}}+805 t^6+\frac{120}{t^6}-25\right) x^2+\left(\frac{10}{t^{17}}+1240 t^7+\frac{126}{t^5}-50 t\right) x^{7/3} \nonumber \\
&\qquad +\left(\frac{1}{t^{40}}+\frac{15}{t^{28}}+\frac{120}{t^{16}}-\frac{24}{t^{10}}+4410 t^8+\frac{680}{t^4}-400 t^2\right) x^{8/3}+\left(\frac{10}{t^{27}}+\frac{126}{t^{15}}+7570 t^9-\frac{45}{t^9}-950 t^3+\frac{855}{t^3}\right) x^3+\cdots,
\end{align}
\normalsize

\noindent where $t$ is a fugacity for the global $U(1)$ symmetry. The $R$-charge of $Q$ is set to be $r_Q=\frac{1}{3}$ for simplicity. The second term $\left(\frac{1}{t^{10}}+15 t^2\right) x^{2/3}$ consists of the meson $M$ and the Coulomb branch operator $Z_{SU(2)}$. The third term $10 t^3 x$ corresponds to the cubic baryon $B$. The quartic baryon $F$ is represented as $5 t^4 x^{4/3}$. The higher order terms are fermion contributions and symmetric products of the bosonic operators. We verified the agreement of the electric and magnetic indices up to $O(x^3)$.

\section{Summary and Discussion}
In this paper, we proposed the Seiberg duality for the 3d $\mathcal{N} = 2$ $Spin(7)$ gauge theory with $F$ spinor matters. The dual description is given by the 3d $\mathcal{N} = 2$ $SU(F-4)$ gauge theory with $F$ anti-fundamental matters, a symmetric tensor and a meson singlet. Since the matter content on the magnetic side is ``chiral'' in a four-dimensional sense, this duality equates ``chiral'' and ``non-chiral'' 3d gauge theories. In Section 4, we switched the roles of the electric and magnetic theories and proposed the 3d Kutasov-type duality for the $SU(N)$ gauge theory with a symmetric tensor and fundamental matters with a (dressed) monopole superpotential. 
As a by-product of the $Spin(7)$ duality, we found the Seiberg duality for the 3d $\mathcal{N}=2$ $G_2$ gauge theory with fundamental matters. These dualities are very similar to the corresponding 4d dualities discovered by Pouliot \cite{Pouliot:1995zc} except for the superpotential. 
An important observation in this paper is that the composite operator $\det \, s$, which was truncated in the 4d duality, is not removed but mapped to the Coulomb branch operator of the dual theory. This is a new feature of the 3d $Spin(7)$ and $G_2$ dualities.   
As a validity check of our analysis, we computed the superconformal indices by using the electric and magnetic theories and observed a beautiful agreement.
 
Although the dualities we found here are very similar to the 4d dualities \cite{Pouliot:1995zc}, we couldn't derive these 3d dualities from the 4d ones via dimensional reduction \cite{Aharony:2013dha, Aharony:2013kma}. An important difference between the 3d and 4d $Spin(7)$ dualities is the absence of the superpotential proportional to $\det \, s$ in 3d. 
It would be interesting if we could find the connection between the 3d and 4d $Spin(7)$ and $G_2$ dualities. In principle, it is possible to reduce the 4d superconformal indices to the 3d partition functions of these dualities \cite{Dolan:2011rp, Gadde:2011ia, Imamura:2011uw, Niarchos:2012ah, Aharony:2013dha}. This will give us another consistency check of the dualities and deepen our understanding. 

The Seiberg dualities proposed here include the (dressed) monopole superpotential. Therefore, this can be regarded as a generalized version of the monopole-deformed dualities \cite{Benini:2017dud}. It is interesting to further generalize the 3d dualities with dressed monopole superpotential in this direction.
Since the theory includes a monopole potential, these dualities have a UV-completion problem. We do not know how to express the dressed Coulomb branch operator in terms of the UV degrees of freedom in a gauge-invariant way. Of course, this does not ruin the validity of the dualities because the proposed dualities are valid in a far-infrared regime. Nonetheless, it is nice if we could understand the origin of the (dressed) monopole superpotential and obtain the UV-complete dualities where both of the electric and magnetic theories are UV-complete and have well-defined Lagrangian descriptions. 

It is very important to generalize our argument to more general cases: One can in principle construct similar dualities for the 3d $\mathcal{N}=2$ $Spin(N)$gauge theories with vector and spinor matters according to the 4d dualities \cite{Pouliot:1995sk, Cho:1997kr, Pouliot:1996zh, Kawano:1996bd, Berkooz:1997bb, Kawano:2007rz, Kawano:2005nc, Cho:1997sa}. We will soon come back to this problem elsewhere.

\section*{Acknowledgments}
This work is supported by the Swiss National Science Foundation (SNF) under grant number PP00P2\_183718/1.


\bibliographystyle{ieeetr}
\bibliography{spin7_duality_ref}

\end{document}